# Imaging gate-tunable Tomonaga-Luttinger liquids in 1H-MoSe$_2$ mirror twin boundaries

Tiancong Zhu[1,2,†], Wei Ruan[1,2,3,†,*], Yan-Qi Wang[1,2,†], Hsin-Zon Tsai[2], Shuopei Wang[4,5], Canxun Zhang[2,6], Tianye Wang[1,2], Franklin Liou[2,6], Kenji Watanabe[7], Takashi Taniguchi[8], Jeffrey B. Neaton[1,2], Alex Weber-Bargioni[1,9], Alex Zettl[2], Ziqiang Qiu[1,2], Guangyu Zhang[4,5,10], Feng Wang[1,2,6,*], Joel E. Moore[1,2,6,*], Michael F. Crommie[1,2,6,*]

[1]*Materials Sciences Division, Lawrence Berkeley National Laboratory, Berkeley, California 94720, USA*

[2]*Department of Physics, University of California, Berkeley, California 94720, USA*

[3]*Department of Physics, Fudan University, Shanghai 200438, China*

[4]*Beijing National Laboratory for Condensed Matter Physics, Key Laboratory for Nanoscale Physics and Devices, Institute of Physics, Chinese Academy of Sciences, Beijing 100190, China*

[5]*Songshan Lake Materials Laboratory, Dongguan, Guangdong 523808, China*

[6]*Kavli Energy Nano Sciences Institute at the University of California Berkeley and the Lawrence Berkeley National Laboratory, Berkeley, California 94720, USA*

[7]*Research Center for Functional Materials, National Institute for Materials Science, 1-1 Namiki, Tsukuba 305-0044, Japan*

[8]*International Center for Materials Nanoarchitectonics, National Institute for Materials Science, 1-1 Namiki, Tsukuba 305-0044, Japan*

[9]*Molecular Foundry, Lawrence Berkeley National Laboratory, California 94720, USA*

[10]*School of Physical Sciences, University of Chinese Academy of Sciences, Beijing 100190, China*

*† These authors contributed equally to this work.*

*\*e-mail: weiruan@fudan.edu.cn, fengwang76@berkeley.edu, jemoore@berkeley.edu, crommie@berkeley.edu*

**Abstract**

One-dimensional electron systems (1DESs) exhibit properties that are fundamentally different from higher-dimensional systems. For example, electron-electron interactions in 1DESs have been predicted to induce Tomonaga-Luttinger liquid behavior. Naturally-occurring grain boundaries in single-layer semiconducting transition metal dichalcogenides provide 1D conducting channels that have been proposed to host Tomonaga-Luttinger liquids, but charge density wave physics has also been suggested to explain their behavior. Clear identification of the electronic ground state of this system has been hampered by an inability to electrostatically gate such boundaries and thereby tune their charge carrier concentration. Here we present a scanning tunneling microscopy/spectroscopy (STM/STS) study of gate-tunable mirror twin boundaries (MTBs) in single-layer 1H-$MoSe_2$ devices. Gating here enables STM spectroscopy to be performed for different MTB electron densities, thus allowing precise characterization of electron-electron interaction effects. Visualization of MTB electronic structure under these conditions allows unambiguous identification of collective density wave excitations having two distinct velocities, in quantitative agreement with the spin-charge separation predicted by finite-length Tomonaga-Luttinger-liquid theory.

Reduced screening, enhanced quantum fluctuations, and confinement effects in one-dimensional electron systems (1DESs) cause a variety of exotic phenomena such as quantum spin liquids[1], Peierls transitions[2], single electron transport[3], and Tomonaga-Luttinger liquids (TLLs)[4-20]. Important characteristics of TLLs include power-law conductance and spin-charge separation, both of which depend sensitively on the ratio between electron-electron (el-el) interaction strength and electron kinetic energy. Direct, precise measurements of the el-el interaction energy in 1DESs, however, have yet to be reported. The power-law conductance[12-14] and spin-charge separation[15,16,19,20] expected for a TLL in the infinite-length limit have been observed in several 1DESs, but the effects of confinement on finite-length 1DESs are not as well understood. Recently discovered mirror twin boundaries (MTBs) in single-layer (SL) 1H-MoX$_2$ (X = S, Se, Te) provide a one-dimensional (1D) metallic channel embedded in a two-dimensional semiconducting bulk that is ideal for investigating such effects over a range of finite lengths[21]. Signatures of an energy gap as well as charge density modulations in MTBs have been experimentally demonstrated[21-27], consistent with theoretical predictions for a finite-length TLL[28-33], but debate exists as to whether their origin is rather due to a Peierls instability[22,25]. Spectroscopic evidence of spin-charge separation in MTBs has been observed[24], but unambiguous identification of the spin/charge excitations and their velocities is still missing.

Here we report the observation of finite-length TLL behavior in MTBs in gate-tunable SL 1H-MoSe$_2$ devices by means of scanning tunneling microscopy/spectroscopy (STM/STS). This work is different from previous measurements because we have incorporated MTBs into gate-tunable devices that allow us to tune the MTB electron-filling while simultaneously performing atomic-scale characterization of MTB electronic structure. This allows us to unambiguously determine the el-el interaction strength in finite MTBs and to compare this to the MTB electron kinetic energy. This was accomplished by measuring the MTB energy gap size as a function of applied gate voltage, as well as observing the MTB charge density distribution, both of which are consistent with expectations for a 1D particle-in-a-box with interaction-induced level-splitting. Clear evidence for TLL-based spin-charge separation in MTBs was obtained through observation of two distinct spectral function modulations that correspond to collective spin and charge excitations with separate dispersions. The characteristic TLL parameter obtained from the observed spin and charge velocities was found to be in excellent agreement with the separately measured MTB el-el interaction energy and energy level spacing.

SL 1H-MoSe$_2$ was grown via molecular beam epitaxy at the surface of epitaxial graphene supported by hBN/SiO$_2$/Si (Fig. 1a, Supplementary Fig. 1). This heterostructure allows systematic control of electron-filling at the MoSe$_2$ surface by way of a doped Si bottom gate (Fig. 1b). The device was characterized by STM topographic imaging (Fig. 1a) which shows large SL MoSe$_2$ islands as well as a moiré superlattice formed by alignment of the graphene and hBN[34]. Isolated MTBs and MTB networks are seen as double-straight-line features that exhibit a 4|4P structure as reported previously[21,22] (Fig. 1c).

The electronic properties of an isolated 14-nm long MTB (outlined in yellow in Fig. 1c) was characterized by measuring bias ($V_b$) dependent STM differential conductance (d$I$/d$V$), which reflects the surface electronic local density of states (LDOS). Figure 1d shows a typical d$I$/d$V$ spectrum at zero gate voltage ($V_g = 0$) that is consistent with previous measurements[22]. Three dominant features in the d$I$/d$V$ spectrum can be identified: an energy gap of $\Delta_0 \approx 128$ meV bracketing the Fermi level ($E_F$) (the peak-to-peak width between v$_0$ and c$_0$), additional peaks further out in energy from the gap (black arrows labeled by v$_1$, v$_2$, c$_1$, c$_2$), and a large peak at $V_b \approx -240$ meV (blue arrow).

The spatial distribution of MTB electronic states at the different peak energies was characterized by performing constant-height d$I$/d$V$ mappings of the MTB in Fig. 1c. Figure 1e shows the electronic LDOS of both the highest occupied state (HOS) at $V_b \approx$ -93 meV ($v_0$) and the lowest unoccupied state (LUS) at $V_b \approx$ 33 meV ($c_0$), revealing periodic charge modulations along the MTB. The HOS is observed to have 13 nodes whereas the LUS has 14, as seen from direct comparison of the LDOS line profiles (Fig. 1f) acquired along the orange and green lines in Fig. 1e (nodes here are defined as local minima in the interior of the MTB). A d$I$/d$V$ map of the $v_1$ peak exhibits 12 nodes while a map of the $c_1$ peak exhibits 15 nodes (Supplementary Fig. 2a-c), suggesting a particle-in-a-box nodal progression (i.e., the number of nodes increases by 1 with each higher energy peak). The peaks $v_i$ and $c_i$ ($i \geqslant 0$) can thus be interpreted as representing confined quantum levels. Similar nodal structure was observed in all the MTBs studied here (see Supplementary Fig. 3 for additional representative data).

To gain insight into the nature of the energy gap at $E_F$ we performed STS measurement of the MTB shown in Fig. 1c for different gate voltages in the range -60 V $\leqslant V_g \leqslant$ 60 V, thus enabling the MTB to be tuned from the hole-doped regime ($V_g$ = -60 V) to the electron-doped regime ($V_g$ = 60 V). Figure 2a shows the gate-dependent d$I$/d$V$ curves acquired at the position marked in Fig. 1e while Fig. 2b shows a d$I$/d$V$ intensity plot for a finer set of gate voltages at the same position. A key observation here is that the gap-size changes with gate-voltage. The d$I$/d$V$ spectrum at $V_g$ = -60 V (Fig. 2a, orange) shows a large gap at $E_F$ of $\Delta_{large} \approx$ 121 meV that is similar to the gap, $\Delta_0$, observed at $V_g$ = 0 V (Fig. 2a, red). The d$I$/d$V$ spectrum at $V_g$ = 60 V (Fig. 2a, blue), however, shows a significantly smaller gap of $\Delta_{small} \approx$ 70 meV. The peaks at the gap-edge of $\Delta_{small}$ are also observed to have reduced intensities compared to those bracketing $\Delta_{large}$. As $V_g$ increases from -60 V the overall bandstructure shifts rigidly towards lower energies, consistent with the electrostatic influence of the bottom gate. For $V_g \approx$ 10 V a pronounced peak (blue arrow) appears at negative bias voltage and shifts to higher energies with increasing $V_g$, opposite to the overall lowering trend seen for the rest of the bandstructure. Such behavior allows us to identify this peak as a tip-induced charging feature of the LUS[35]. For $V_g \approx 20$ V both the charging peak and the LUS begin to cross $E_F$, resulting in a transition from the large energy gap to the smaller gap. For $V_g > 20$ V both the charging peak and the overall bandstructure continues to shift as expected for increased electron filling of the MTB. Similar gate-dependent behavior was also observed for 19 other MTBs that were similarly measured using a variety of different STM tips (see Supplementary Fig. 4 for representative data).

Constant-height d$I$/d$V$ maps of the HOS and LUS of the MTB in Fig. 1c at $V_g = \pm 60$ V further reveal gate-dependent real-space electronic structure (Fig. 3). The HOS and LUS for the large gap configuration at $V_g = -60$ V ($\Delta_{\text{large}}$) exhibit 13 nodes and 14 nodes, respectively (Fig. 3b), consistent with the real-space electronic structure observed for the undoped ($V_g = 0$ V) case (Fig. 1e, f). The HOS and LUS for the small gap configuration at $V_g = 60$ V ($\Delta_{\text{small}}$), on the other hand, both exhibit 14 nodes (Fig. 3e). The LDOS maps of higher energy peaks are summarized in Supplementary Fig. 2d-i. (Similar gate-dependent nodal structure was observed for 9 MTBs).

In order to characterize the energy- *and* momentum-resolved MTB electronic structure we measured d$I$/d$V$ spectra along an MTB for the large gap case ($V_g = 0$V) (Fig. 4a) and performed Fourier transform (FT) analysis of the resulting density plot (here we chose a longer MTB than that in Figs. 1-3 to achieve better momentum resolution). Figure 4b shows the energy-dependence of the STS intensity plot as a function of the MTB axial coordinate ($x$ axis) and the sample bias ($y$ axis). Fast real-space modulations ($\lambda \sim 1$ nm) create a complex MTB nodal structure that coexists with a longer wavelength modulation that induces a dome-shaped charge density profile (similar real-space modulations in electronic structure have been previously reported both for MTBs[19,21] and carbon nanotubes[15,16]). Figure 4c shows the corresponding FT of the STS intensity plot, revealing both the energy- and momentum-dependence of the electronic structure. Two linear dispersion branches with different slopes are seen to cross $E_F$ at $q/2\pi \approx 1$ nm$^{-1}$ (which we identify as $q = 2k_F$ where $k_F$ is the Fermi wavevector) and are labeled by blue and red markers in Fig. 4c. These branches correspond to "fast" real-space nodal structure and are dubbed the "fast branches" hereafter. The velocity of the blue branch is 3.5×10$^5$ m/s (extracted from twice the slope of the dispersion since LDOS $\propto |\psi(x)|^2$) and is consistent with the Fermi velocity, $v_F$, of the MTB metallic bandstructure[22,24]. The velocity of the red branch, on the other hand, is 6.5×10$^5$ m/s and is significantly higher than $v_F$. An additional linear branch near the Γ point (labeled by orange markers in Fig. 4c) can be resolved that corresponds to the long wavelength modulation mentioned above (and is dubbed the "slow branch" hereafter). The existence of multiple linear dispersion branches causes blurring of the real-space nodal structure at peak energies far from $E_F$ (Supplementary Fig. 3). A static mode (i.e., constant wavevector with energy) at $q = 2k_F$ can also be observed, consistent with previous results[22]. Fig. 4d provides a 2$^{nd}$ derivative plot of the data presented in Fig. 4c which allows the various dispersive features to be more clearly seen. This type of behavior was observed in all 7 of the MTBs that were characterized in this way (see Supplementary Fig. 5 for additional representative data). All the measured MTBs were well isolated from other MTBs and defects to avoid spatial inhomogeneities in the local electrostatic environment that might obscure the intrinsic MTB behavior.

    Our experimental results are in excellent agreement with the theoretical predictions for a confined TLL. Our observation of gate-induced modulation of the energy gap size reveals the strength of el-el Coulomb repulsion in a MTB and thus confirms that they should, indeed, be described by TLL theory. Were el-el interactions absent then the Fermi-level energy gap would be gate-independent and would simply equal the 1D single-particle level-spacing value of $E_0 = \hbar v_F \pi / L$. The presence of el-el Coulomb repulsion explains the large and small gap variation seen for different electron fillings since each MTB quantum confinement

level has spin degeneracy and can accept two electrons. When one of the spin states is filled by tuning the gate voltage then el-el repulsion will create a charging gap $\Delta_{\text{small}} = E_C$ that is required to inject a second electron with the opposite spin into the same MTB level, resulting in level splitting and a magnetic ground state (Fig. 3f, green). This explains the identical real-space nodal structure experimentally observed for the HOS and LUS in Fig. 3e, as well as the reduced spectral weight of the HOS/LUS states in Fig. 2 at $V_g = 60$ V (where the HOS and LUS both contain just a single spin state) compared to $V_g = -60$ V (where the HOS and LUS are both spin degenerate). When the MTB HOS level contains two electrons (e.g., at $V_g = -60$ V) then repulsion between the HOS and LUS electrons adds $E_C$ to the single-particle level-spacing, $E_0$, resulting in a larger gap $\Delta_{\text{large}} = E_0 + E_C$ (Fig. 3c). In this case we expect a different number of nodes for the HOS and LUS since they are associated with different quantum confinement levels, as seen experimentally in Fig. 3b. This scenario is also supported by the fact that the single-particle level-spacing defined by the energy difference between $v_1$ and the HOS (e.g., $E_0 = 52$ meV (Fig. 1d)) matches the value estimated from a different set of levels by $E_0 = \Delta_{\text{large}} - \Delta_{\text{small}}$ (e.g., 51 meV (Fig. 2a)).

Our observation of gate-induced modulation of the MTB energy gap allows a direct and precise measurement of $E_C/E_0$, the ratio of MTB el-el interaction energy to electron kinetic energy. Since the single-particle level-spacing can be expressed as $E_0 = \hbar v_F \pi / L$ and the charging energy $E_C$ is inversely proportional to the separation of two electrons, $\langle x_1 - x_2 \rangle \sim L$, we expect both $\Delta_{\text{large}}$ and $\Delta_{\text{small}}$ to scale as $L^{-1}$. This is confirmed by measurements of the gap-size statistics for MTBs of different lengths ranging from 6 nm to 30 nm (Fig. 5a). The ratio of $E_C/E_0$ is thus observed to be universal for MTBs of different lengths. This ratio is related to TLL behavior through the TLL parameter $K_c$ as[12,31]

$$K_c = \left(1 + \frac{2E_C}{E_0}\right)^{-\frac{1}{2}} \tag{1}$$

(see Supplementary Note 1.1). The energy gaps measured in our spectroscopy of MTBs having different lengths yield a universal value of $K_c = 0.54 \pm 0.03$ (Fig. 5b).

Another piece of evidence supporting the TLL interpretation of MTBs is the observation of spin-charge separation. The existence of two linear fast branches in the FT-STS suggests that a single metallic band picture[22,24] for MTBs is insufficient, whereas TLL-based spin-charge separation can explain this quantitatively. To see this we performed simulations of the expected LDOS for a finite 1D TLL[24,32] having $K_c = 0.54$ (Fig. 4e-g) and with all other material-dependent parameters constrained by experimental values (see Supplementary Note 1.2). Both the fast and slow modulations can be seen in the resulting theoretical energy-dependent LDOS (Fig. 4e), which closely resembles the experimental features seen in Fig. 4b. Clear spin-charge separation can be observed in the FT of the simulated LDOS, as seen in Fig. 4f where two linear branches show LDOS modulations induced by spin (blue arrow) and charge (red arrow) density excitations that have distinct velocities[24,28,32,33] (the slow branch and static $2k_F$ branch can also both be seen). Direct comparison between Fig. 4c and Fig. 4f allows us to identify the blue and red branches in the experimental data of Fig. 4c as the TLL spin and charge branches, respectively (this can be seen even better by comparing the second derivative dispersions in Figs. 4d and 4g). The clear signature of spin-charge separation seen in the FT-STS measurement allows us to experimentally extract the TLL parameter through a second, independent method via the following relation:[8,9,15,16,29,30,36]

$$K_c = \frac{v_s}{v_c}, \qquad (2)$$

where $v_s$ and $v_c$ are the slopes (i.e., velocities) of the spin and charge branches, respectively (Supplementary Note 1.1). This technique yields a value of $K_c = 0.53 \pm 0.05$ (Fig. 5c), in excellent agreement with the value of $K_c = 0.54 \pm 0.03$ determined from the energy gap ratio $E_C/E_0$ (Fig. 5b). The TLL picture is thus confirmed through self-consistent measurements of both energy level alignment (Fig. 2) and spatial LDOS modulations (Fig. 4).

In conclusion, our observation of gate-dependent energy gaps via STS allows precise measurement of the el-el Coulomb interaction energy in 1D MTBs. Spin-charge separation visualized via FT-STS allows us to confirm that this interaction indeed leads to TLL behavior, including distinct spin and charge velocities. The TLL parameter determined separately from the el-el interaction energy and spin/charge velocities are both in excellent agreement with each other. Our experiment demonstrates that combining STM with gate-tunable single-layer devices is a useful approach for gaining a better understanding of electron correlation effects in nanoscale systems. Our device thus provides an ideal platform to study the response of 1D TLL systems to magnetic scatterers[37,38], external magnetic field[39], and tuned dielectric environments[26].

**Acknowledgments**

This research was supported as part of the Center for Novel Pathways to Quantum Coherence in Materials, an Energy Frontier Research Center funded by the U.S. Department of Energy, Office of Science, Basic Energy Sciences (material growth, STM spectroscopy, theoretical simulations). Support was also provided by the National Science Foundation through grant DMR-1807233 (device design). S.W. and G.Z. acknowledge support by Guangdong Basic and Applied Basic Research Foundation through grant No. 2019A1515110898 (epitaxial graphene growth). Z.Q. acknowledges support by the National Research Foundation of Korea through grant No. 2015M3D1A1070467 (MBE instrumentation development) and No. 2015R1A5A1009962 (MBE growth characterization). K.W. and T.T. acknowledge support from the Elemental Strategy Initiative conducted by the MEXT, Japan, Grant Number JPMXP0112101001 (hBN growth), JSPS KAKENHI Grant Numbers 19H05790 (hBN chacterization) and JP20H00354 (development of new hBN growth tools).

**Methods**

**Sample fabrication**

The preparation of the epitaxial graphene/hBN heterostructure supported on $SiO_2$/Si substrate is described in ref.[34]. Single-layer 1H-$MoSe_2$ films with MTBs were grown directly on epitaxial graphene/hBN heterostructures using MBE. Mo and Se were evaporated from an e-beam evaporator and a home-built Knudsen cell, respectively. The flux ratio between Mo and Se was ~ 1:100 and the sample was kept at 450 ºC during the growth. After growth the sample was heated up to ~600 ºC and annealed under Se for 30 min. The sample was capped with ~20 nm of amorphous Se before being taken out of the ultrahigh-vaccum (UHV) growth chamber. Electrical contacts were made by depositing Cr/Au (3 nm/30 nm) through a shadow mask.

**STM/STS measurements**

Scanning tunneling miscroscopy/spectroscopy (STM/STS) measurements were performed in a low-temperature UHV STM system (CreaTec) at $T$ = 5 K. Prior to measurements the samples were annealed in UHV at ~200 °C for 1 hour to remove the Se capping layers and then immediately transferred in situ to the STM stage at $T$ = 5 K. Electrochemically etched tungsten tips were calibrated on a Au(111) surface before other measurements. d$I$/d$V$ spectra were collected using standard lock-in techniques ($f$ = 401 Hz). d$I$/d$V$ mapping was performed in constant-height mode (i.e., with the feedback loop open).

**Data availability:**

The data that support the findings of this study are available from the corresponding authors upon reasonable request.

**Code availability:**

The codes used in this study are available from the corresponding author upon reasonable request.

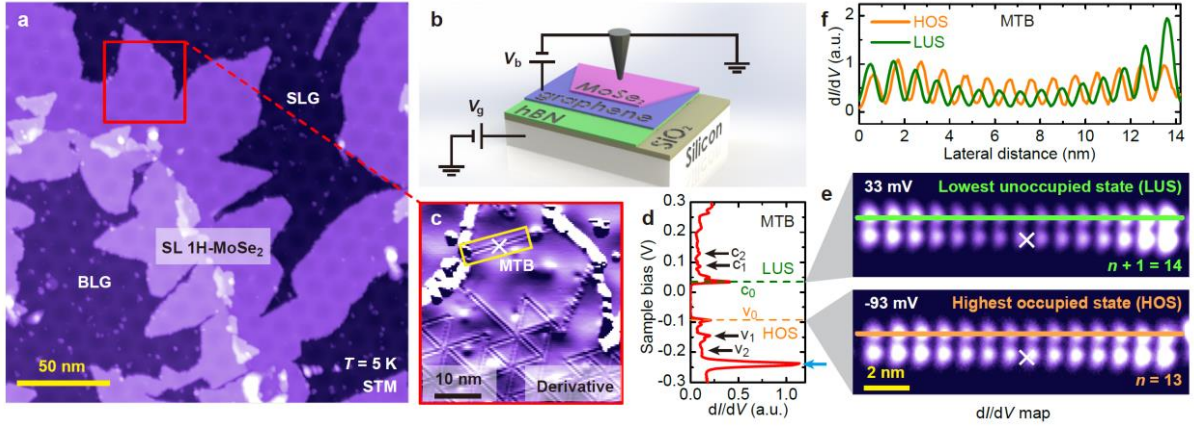

**Fig. 1. STM characterization of a SL MoSe$_2$/Graphene/hBN/SiO$_2$/Si device at *T* = 5 K.**
**a**, Large-scale STM topography of a single-layer MoSe$_2$/Graphene/hBN/SiO$_2$/Si device ($V_b$ = -2 V, $I_t$ = 10 pA). **b**, Schematic of the experimental device setup. **c**, A close-up STM derivative image of the mirror twin boundaries (MTBs) in the area indicated by a red square in **a** (the derivative plot enhances the MTB image contrast). **d**, d$I$/d$V$ spectrum acquired on the MTB in the yellow rectangle in **c** at the position marked by a white cross ($V_b$ = -0.3 V, $I_t$ = 100 pA, $V_{mod}$ = 2 mV). The energy gap is bracketed by the highest occupied state (HOS, $v_0$) and the lowest unoccupied state (LUS, $c_0$). Other 1D quantum well states are labeled with black arrows while a charging peak is labeled with a blue arrow. **e**, Constant-height d$I$/d$V$ maps of the MTB in **c** taken at the LUS (top) and the HOS (bottom) ($V_{mod}$ = 10 mV). The HOS map exhibits 13 nodes while the LUS map exhibits 14 nodes. **f**, LDOS line profiles of the MTB at the energies of both the LUS and the HOS, acquired along the orange and the green lines in **e**.

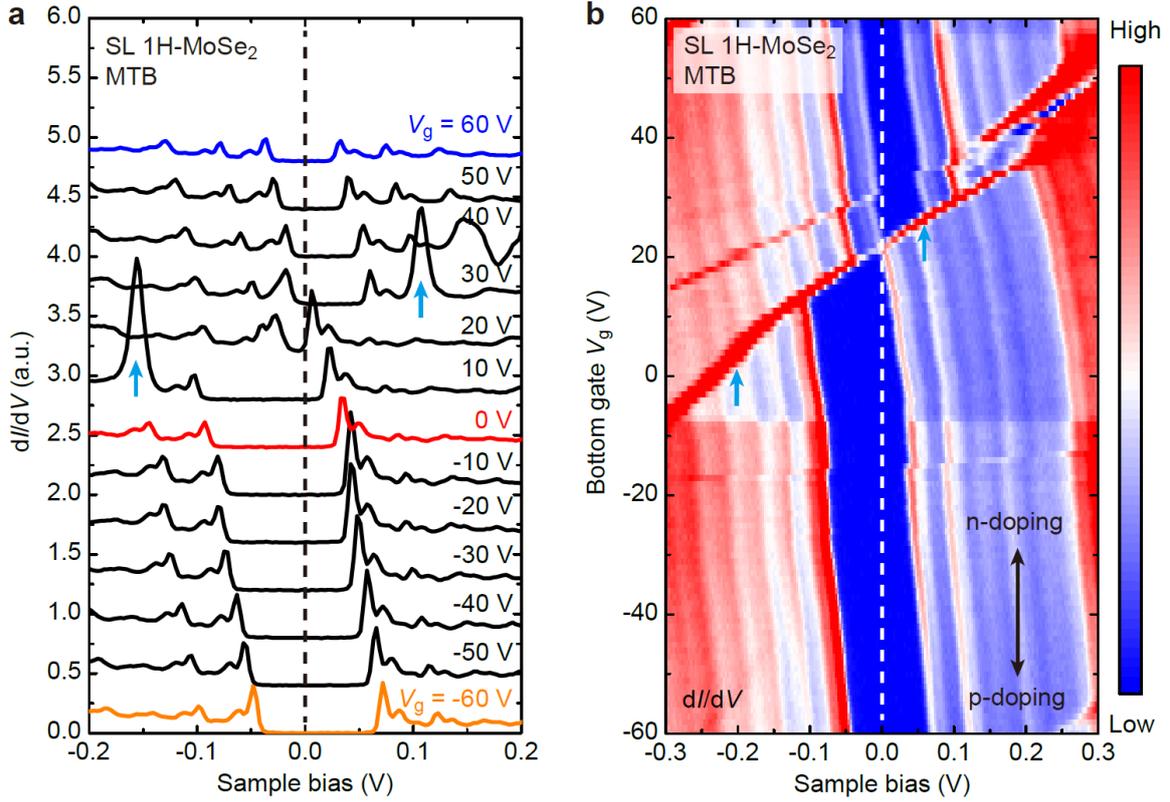

**Fig. 2. Gate dependent electronic structure of the MTB.**

**a**, A waterfall plot showing d$I$/d$V$ spectra acquired over the gate voltage range -60 V ⩽ $V_g$ ⩽ 60 V on the MTB from Fig. 1c ($V_b$ = -0.3 V, $I_t$ = 100 pA, $V_{mod}$ = 2 mV, position marked in Fig. 1c). The spectra exhibit a large energy gap for $V_g$ < 20 V and a small gap for $V_g$ > 20 V. Electronic structure shifts to lower energy with increasing $V_g$ (i.e., with increased electron doping). **b**, A density plot of d$I$/d$V$ spectra acquired for -60 V ⩽ $V_g$ ⩽ 60 V on the MTB showing the overall bandstructure shift and energy gap transition as a function of $V_g$ ($V_b$ = -0.3 V, $I_t$ = 100 pA, $V_{mod}$ = 5 mV).

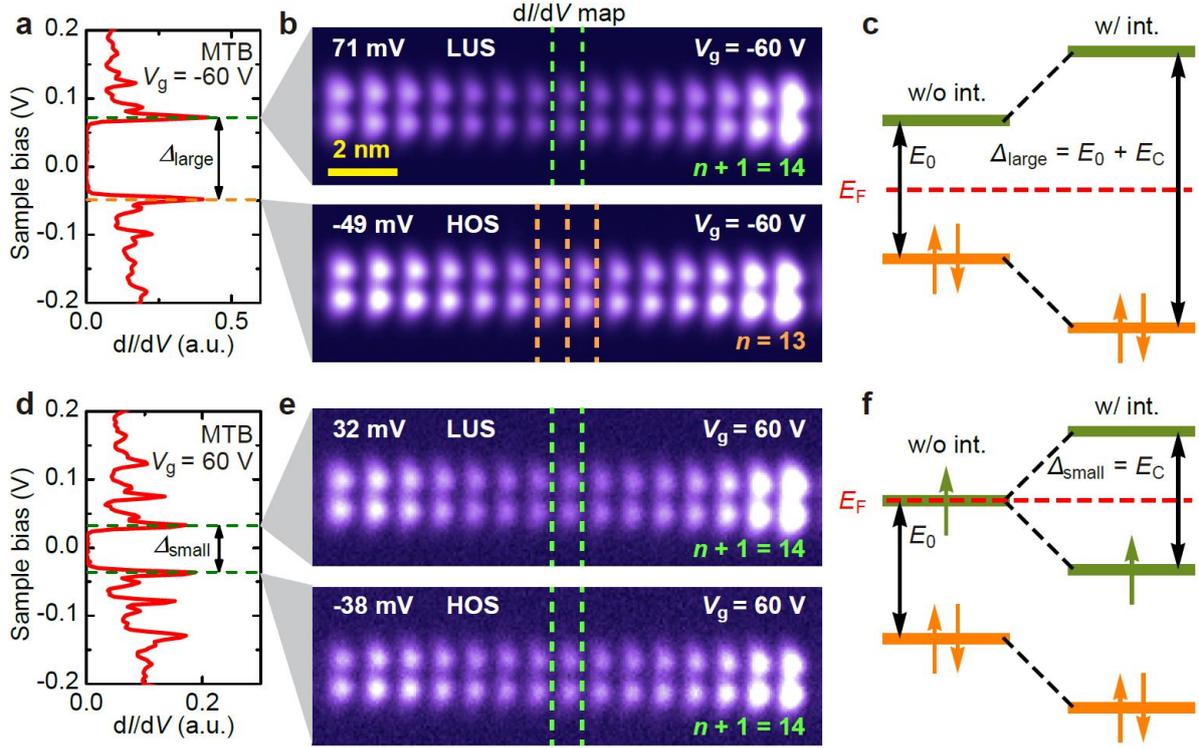

**Fig. 3. Electronic LDOS maps of states at gap edges for $V_g = -60$ V and $V_g = 60$ V.**

**a**, d$I$/d$V$ spectrum of the MTB in Fig. 1c taken at $V_g = -60$ V shows a large energy gap ($V_b = -0.3$ V, $I_t = 100$ pA, $V_{mod} = 2$ mV, position marked in Fig. 1c). **b**, Constant-height d$I$/d$V$ maps of states at gap edges for $V_g = -60$ V ($V_{mod} = 10$ mV). The number of nodes for the HOS and the LUS differ by one. **c**, Energy level diagram for the large energy gap case with and without el-el interactions. **d**, d$I$/d$V$ spectrum of the same MTB taken at $V_g = 60$ V shows a small energy gap ($V_b = -0.3$ V, $I_t = 100$ pA, $V_{mod} = 2$ mV, same position). **e**, Constant-height d$I$/d$V$ maps of states at the gap edges for $V_g = 60$ V ($V_{mod} = 10$ mV). The number of nodes for the HOS and the LUS are now the same. **f**, Energy level diagram for the small energy gap case with and without el-el interactions.

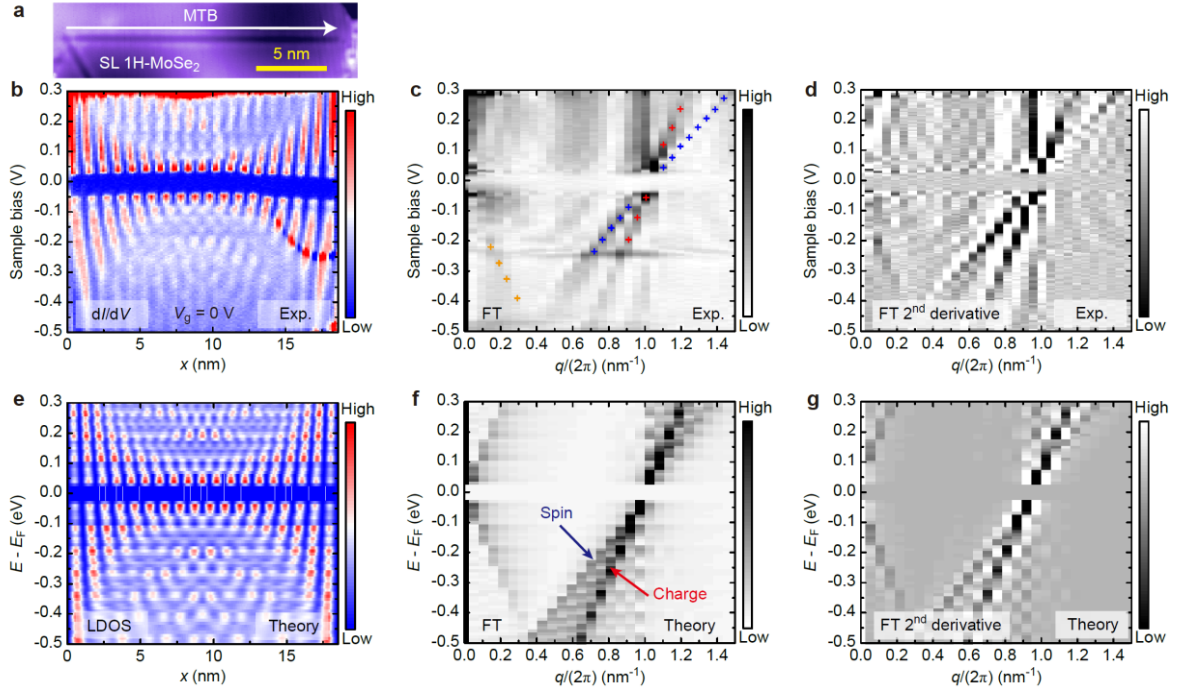

**Fig. 4. Experimental STS along MTB at $V_g = 0$ V compared to theoretical LDOS based on TLL model.**

**a**, STM topography of an 18-nm long MTB ($V_b = -2$ V, $I_t = 10$ pA). **b**, d$I$/d$V$ intensity plot of the MTB along the white arrow in **a** as a function of sample bias $V_b$ ($V_{mod} = 5$ mV, tip positioning parameters: $V_b = -0.6$ V, $I_t = 100$ pA). **c**, Fourier transform (FT) of the d$I$/d$V$ data in **b** as a function of $V_b$ and wavevector $q$. Main features include two linear fast branches with different velocities (marked blue and red), a slow branch near the Γ point (orange dashed line), and a static branch with energy-independent wavevector at $q/2\pi \approx 1$ nm$^{-1}$ ($= 2k_F/2\pi$). **d**, Second-derivative plot of FT in **c** enhances image contrast of the spectral features. **e**, Theoretical LDOS predicted by the finite TLL model using $E_{gap} = 0.08$ eV, $K_c = 0.54$, $K_s = 1$, $k_F^+ = 19\pi/L$, $k_F^- = 18\pi/L$, $v_c\pi/L = 0.073$ eV, and $v_s\pi/L = 0.039$ eV. **f**, FT of the theoretical LDOS in **e** shows similar dispersive and static branches as the experimental data in **c**. **g**, Second-derivative plot of the theoretical FT in **f** for comparison to experimental data in **d**.

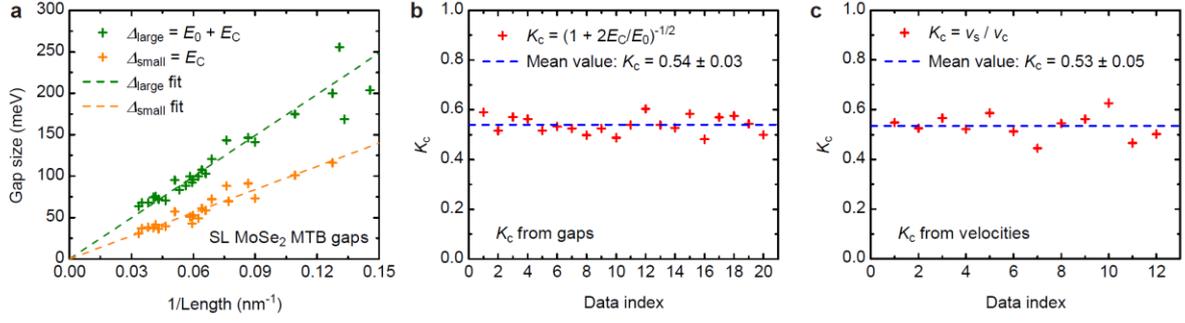

**Fig. 5. Gap-size statistics and MTB TLL parameter obtained in two different ways.**

**a**, Measured large (green) and small (orange) energy gaps for MTBs of different lengths. Both types of gap scale in size as $1/L$ where $L$ is the MTB length. The dashed lines represent linear fits to the data. **b**, The TLL parameter obtained from the measured charging energy, $E_C$, and single-particle level spacing, $E_0$, for MTBs having different lengths ($K_c = 0.54 \pm 0.03$). **c**, The TLL parameter obtained from the ratio of the measured spin and charge velocities for MTBs having different lengths ($K_c = 0.53 \pm 0.05$). The excellent agreement between the $K_c$ values obtained using these two different experimental techniques provides strong evidence for TLL behavior in SL MoSe$_2$ MTBs.